\documentclass[12pt]{JHEP3}
\usepackage{amsfonts,amssymb,amsmath}

\title{A note on the universality of the Hagedorn behavior of pp-wave strings}

\author{Alex Hamilton\footnote{ahamil@gmail.com}, Jeff Murugan\footnote{jeff@nassp.uct.ac.za} and Andrea Prinsloo\footnote{prnand004@mail.uct.ac.za} \\
Cosmology and Gravity Group,\\ 
Department of Mathematics and Applied Mathematics, \\
University of Cape Town, \\ 
Private Bag, Rondebosch, 7700 \\
South Africa}

\abstract{Following on from recent studies of string theory on a one-parameter family of integrable deformations of $AdS_{5}\times S^{5}$ proposed by Lunin and Maldacena, we carry out a systematic analysis of the high temperature properties of type IIB strings on the associated pp-wave geometries. In particular, through the computation of the thermal partition function and free energy we find that not only does the theory exhibit a Hagedorn transition in both the $(J,0,0)$ and $(J,J,J)$ class of pp-waves, but that the Hagedorn temperature is insensitive to the deformation suggesting an interesting universality in the high temperature behaviour of the pp-wave string theory. We comment also on the implications of this universality on the confinement/deconfinement transition in the dual $\mathcal{N}=1$ Leigh-Strassler deformation of ${\cal N}=4$ Yang-Mills theory.}
\preprint{}

\begin{document}

\newcommand{\pl}{\partial}
\newcommand{\be}{\begin{equation}}
\newcommand{\ee}{\end{equation}}
\newcommand{\bea}{\begin{eqnarray}}
\newcommand{\eea}{\end{eqnarray}}
\newcommand{\beas}{\begin{eqnarray*}}
\newcommand{\eeas}{\end{eqnarray*}}
\parskip 10 pt
\parindent 0 pt

 \section{Introduction}

In the ten years since its discovery, the AdS/CFT correspondence has established 
itself as one of the cornerstones of contemporary high energy physics 
\cite{Maldacena97}.   On one hand, this correspondence is arguably the first - albeit only conjectural -
non-perturbative realization of string theory. Consequently, it is anticipated that a complete understanding of the nature of the correspondence will yield many of the mysteries that shroud the 
regime of quantum gravity.  On the other hand, as a strong/weak coupling duality, the gauge theory/gravity correspondence as embodied in Maldacena's conjecture has proven an invaluable tool in the understanding of the non-perturbative sector of gauge theory with applications everywhere from the
computations of gluon scattering amplitudes in ${\cal N}=4$ at strong coupling \cite{Alday-Maldacena07} to a conjectured universal lower bound on the shear-viscosity to entropy density 
ratio of the strongly coupled quark-gluon plasma produced at RHIC \cite{PSS01}.

However, even after many such remarkable discoveries and ten years of development, 
the AdS/CFT conjecture remains just that; a conjecture. This, as with many of the foundational
problems of string theory, is attributable in no small part to a fundamental lack of understanding of 
the complete set of degrees of freedom of the theory. While strings provide a useful accounting of the perturbative states of the theory, the discovery of D-branes \cite{Polchinski95}
revealed - very pointedly so - that they were by no means the only states. An excellent exemplification 
is found in the high energy growth of point gravitons - closed string KK-modes - into giant gravitons 
described in the type IIB supergravity by spherical D3-branes \cite{MST00}. Actually, the study of finite temperature effects in string theory, most notably the exponential growth of states near the Hagedorn temperature and the consequent divergence of the canonical partition function, reveals that strings may not even be the fundamental constituents of the theory \cite{Hagedorn-Refs}. Supposedly, just as the partonic structure of QCD emerges at high temperatures and energies, so too will the true fundamental degrees of freedom of string theory. To claim a complete understanding of the behavior of string theory above the Hagedorn temperature is thus a gross understatement.

On the gauge theory side of things, while it has long been suspected that the Hagedorn behavior of string theory on $AdS_{5}\times S^{5}$ is intimately related to the confinement/deconfinement transition in planar ${\cal N}=4$ SYM on ${\mathbb R}\times S^{3}$ (see for instance \cite{SYM Hagedorn refs}), matching the temperatures on both sides of the duality is no mean feat. In general, a direct quantitative computation of the temperature (as well as the behavior of strings near the Hagedorn point) is hampered by the need for an explicit, quantized string spectrum - something decidedly lacking in the case of the $AdS_{5}\times S^{5}$ superstring. Consequently, most of the literature on the field is restricted to flat backgrounds or toroidal compactifications thereof (as in, for example, the second of \cite{Hagedorn-Refs}). Fortunately, there {\it is} another entirely non-trivial arena in which to test the correspondence: the maximally symmetric pp-wave background obtained from $AdS_{5}\times S^{5}$ by a suitable Penrose scaling limit \cite{BMN}. Not only is string theory on this background exactly soluble (and hence yielding of an exact quantum string spectrum) \cite{Metsaev:2001bj} but through a remarkable insight of Berenstein, Maldacena and Nastase in \cite{BMN}, the dual gauge theory (or, more accurately, the specific sector of ${\cal N}=4$ SYM) is also explicitly known. The study of thermal strings on this background that followed shortly thereafter \cite{pp-wave Hagedorn} proved to be extremely fruitful, demonstrating that not only does the Hagedorn temperature (and the accompanying exponential growth of states) exist in this background but that it is also an indication of a phase transition rather than a hard cutoff to the temperature. This would seem to mesh neatly with the confinement/deconfinement transition observed in the planar sector of the Yang-Mills theory except for the fact that in the BMN double scaling limit, only a subsector of states survive - the ``near BPS'' states, whose anomalous dimensions lie systematically close to chiral primaries.

The problem is again one of compatibility of regimes; where we are finally able to compute an exact string spectrum (and consequently, the Hagedorn temperature) on the gravity side of the duality, there are an insufficient number of states in the dual gauge theory sector to account for the required exponential growth. Circumventing this difficulty is quite nontrivial and was only recently accomplished by identifying an altogether different decoupling limit  of thermal $SU(N)$,  ${\cal N}=4$ SYM in which the physics of the gauge theory is exactly captured by a ferromagnetic XXX$\displaystyle{}_{\frac{1}{2}}$ Heisenberg spin chain \cite{Matching Hagedorns}. The Hagedorn temperature is then computed from well-known thermodynamic properties of the spin chain and matches excellently with the string theory result. 

If, as it is hoped, the AdS/CFT program is to eventually lead to a complete understanding of QCD - a decidedly non-conformal, non-supersymmetric and non-Abelian theory with very finite N - it is crucial to identify among the plethora of remarkable results known for ${\cal N}=4$ SYM which are a consequence of the large amounts of symmetry and which are truly universal. It is with this in mind that we investigate in this article the possibility that the Hagedorn temperature of the pp-wave string might be just such a universal property of the theory. Of particular interest to us is the Hagedorn behavior of the pp-wave string under a systematic breaking of supersymmetry of the sort that comes with a recently discovered integrable deformation of $AdS_{5}\times S^{5}$ \cite{LM}. This Lunin-Maldacena transformation generates a one parameter family of marginal deformations of AdS preserving an ${\cal N}=1$ supersymmetry and exactly dual to the Leigh-Strassler superpotential deformation of ${\cal N}=4$ SYM \cite{Leigh-Strassler}. Unlike the original geometry in which all Penrose limits lead to the same pp-wave background (essentially due to the $SO(6)$ isometry of the 5-sphere), the Lunin-Maldacena deformation $AdS_{5}\times S^{5}_{\gamma}$ supports two inequivalent Penrose limits, distinguished by choice of null geodesic. String theory on both plane wave backgrounds has been studied in some detail with rather remarkable results. In particular, taking the Penrose limit about any one of the $(J,0,0), (0,J,0)$ or $(0,0,J)$ class of geodesics yields a conventional pp-wave background on which the string spectrum exhibits a distinct dependence on the deformation parameter \cite{LM,Niarchos-Prezas}. 
The second set of BPS geometries, resulting from taking the Penrose limit about null geodesics with angular momenta $(J,J,J)$, is a set of homogeneous plane waves whose metric lies in a different diffeomorphism class to that of the former \cite{Hpp}. Intriguingly, the spectrum of strings on this spacetime is independent of the deformation parameter\footnote{This is a result of a cunning conspiracy between the NS $B-$field turned on by the Lunin-Maldacena transformation and the metric deformation. See \cite{Deformed giants} for more discussion on this point.} - a result verified to one loop in the spectrum of anomalous dimensions of near-BPS operators in the dual ${\cal N}=1$ gauge theory \cite{DMSS,Mateos}.   

Taking these spectra as a launching point, this article is organized as follows: In the interests of self-containment and to establish our notation, in the following section we give a detailed derivation of the single- and multi-string partition functions for strings on the (undeformed) pp-wave background pointing out several subtleties along the way that will factor into our later computations. With this foundation in place we present in section 3 a very brief outline of the Lunin-Maldacena deformation, its various BPS pp-wave limits and the oscillation spectra of strings on these backgrounds. Restricting our attention to  the $(J,0,0)$ case, section 3.3 contains our derivation of the deformed multi-string partition function and corresponding Hagedorn temperature respectively. In section 4 we discuss, using the recent prescription of \cite{Matching Hagedorns}, the matching of the Hagedorn temperature of the deformed pp-wave string to the confinement/deconfinement transition of the dual gauge theory.  Finally, our discussions and speculations on future directions are contained in the Conclusion. 

\section{Thermodynamics of pp-wave strings} \label{SUGRA}

In this section, we present a detailed review of the thermodynamics of strings
moving in a maximally supersymmetric pp-wave background, closely
following discussions in the second reference of \cite{pp-wave Hagedorn}. This pp-wave background
is described by the metric
\begin{equation}
ds^{2} = -2dx^{+}dx^{-} - \mu^{2}\sum_{i=1}^{8}\left(x^{i}\right)^{2}\left(dx^{+}\right)^{2} + \sum_{i=1}^{8}\left(dx^{i}\right)^{2},
\end{equation}
where $x^{\pm}$ are the light-cone coordinates, and the $x^{i}$
describe eight transverse directions.  Amongst the numerous symmetries
contained in this metric are an $SO(8)$ rotational symmetry in the
transverse coordinates (broken to $SO(4) \times SO(4)$ by the
five-form field), 16 boost-like symmetries in the
$x^{i},x^{-}$-planes, and two translational isometries in the
light-cone $x^{\pm}$ directions - it is these isometries which are
essential for the construction of the single pp-wave string partition
function.

\subsection{Single string partition function}

The partition function describing a single pp-wave string (in the
canonical ensemble) moving in a heat bath at temperature $T$ can be
constructed using a combination of the two
translational isometries as follows:
\begin{equation}
Z_{1}(a,b) = \textrm{Tr}_{\mathcal{H}}~e^{-ap_{+}-bp_{-}},
\end{equation}
Here $p_{\pm} \sim -i\partial_{\pm}$, while the two variables $a$ and $b$ determine the heat bath temperature $T$, which satisfies
\begin{equation}
T^{-2}=ab+a^{2}\mu^{2}\sum\limits_{i=1}^{8}\left(x^{i}\right)^{2}.
\end{equation}

We shall now consider the pp-wave string in the light-cone
gauge\footnote{The light-cone gauge is obtained by making use of the
  conformal gauge, in which the worldsheet metric is diagonal,
  $h_{\alpha\beta}\propto\textrm{diag}(-1,+1)$, and then setting
  $X^{+}(\tau,\sigma)=X^{+}_{0} + (2\alpha'p^{+})\tau$, with $p^{+}$ a
  fixed constant (see \cite{plefka} for further details).}, in which
$p_{-}$ is fixed, while $p_{+}$, the light-cone Hamiltonian is
\begin{equation}
H_{\mathrm{LC}} = 
\frac{1}{2 \alpha' p_{-}}\left[ \omega_{0}\left(N^{B}_{0} + N^{F}_{0}\right) + 
\sum_{n \geq 1}\omega_{n}\left(N^{B}_{n} + N^{F}_{n} + 
\tilde{N}^{B}_{n} + \tilde{N}^{F}_{n}\right) \right],
\end{equation}
where $\omega_{n} \equiv \textrm{sign}(n)\sqrt{n^{2}+m^{2}}$, with
$m=2\mu\alpha'p_{-}$, and $N^{B,F}_{n}$ and $\tilde{N}^{B,F}_{n}$ are
the right- and left-moving number operators describing the eight
bosonic and eight fermionic modes. (The right- and left-moving zero
modes have been identified.) The zero point energy cancels out due to
the supersymmetry.  The level-matching constraint
\begin{equation}
\mathcal{P} = \sum_{n\geq 1} n\left(N^{B}_{n} + N^{F}_{n} - \tilde{N}^{B}_{n} - \tilde{N}^{F}_{n}\right) = 0,
\end{equation}
which arises as a result of worldsheet translation invariance, must
also be satisfied.

The single string partition function may now be written in the form
\begin{equation}\label{single string partition function}
Z_{1}(a,b,m) = \int_{0}^{\infty}dp^{+}\int_{-\frac{1}{2}}^{+\frac{1}{2}} d\tau_{1}~ e^{-bp^{+}}~z_{lc}\left(\tau_{1},\frac{a}{4\pi\alpha'p^{+}},m\right),
\end{equation}
with
\begin{equation}
z_{lc}(\tau_{1},\tau_{2},m) \equiv \textrm{Tr}_{\rm states}
~e^{-2\pi\tau_{2}H_{\mathrm{WS}} + 2\pi i\tau_{1}\mathcal{P}}.
\end{equation}
The trace runs over all the eigenstates of the worldsheet Hamiltonian ($H_{WS} = 2\alpha' p_-$).
The level-matching constraint is imposed using the delta function,
which arises from the integral over $\tau_{1}$.

Finally, it is known that this single string partition function may be
written in terms of building blocks $\Theta_{\alpha,\delta}$. More
specifically, we find that\footnote{The numerator and denominator of
  this ratio of building blocks describe the contributions from the
  fermionic and bosonic modes respectively.},
\begin{equation}\label{zlc-theta}
z_{lc}(\tau_{1},\tau_{2},m) = \left[\frac{\Theta_{\frac{1}{2},0}(\tau_{1},\tau_{2},m)}{\Theta_{0,0}(\tau_{1},\tau_{2},m)}\right]^{4},
\end{equation}
with\footnote{The two terms in the product describe two fields, which
  are complex conjugates of each other, while the left- and
  right-moving modes are captured by negative and positive values of
  $n$ respectively.}
\begin{eqnarray}\label{building blocks - def}
\nonumber & \Theta_{\alpha,\delta}(\tau_{1},\tau_{2},m) & \equiv e^{4\pi\tau_{2}E_{\delta}(m)}\prod_{n=-\infty}^{\infty}
\left(1 - e^{-2\pi\tau_{2}|\omega_{n+\delta}| + 2\pi i\tau_{1}(n+\delta) + 2\pi i \alpha}\right) \\
&& ~~~~~~~~~~ ~~~~~~~~~~ ~~ \times \left(1 - e^{-2\pi\tau_{2}|\omega_{n-\delta}| + 2\pi i\tau_{1}(n-\delta) - 2\pi i \alpha}\right).~~~~~~
\end{eqnarray}
Here $E_{\delta}(m)$ is the casimir energy of a complex boson
of mass $m$ with boundary conditions $\phi(\sigma+2\pi,\tau) = e^{2\pi
  i\delta}\phi(\sigma,\tau)$ \cite{Takayanagi:2002pi}. This casimir
energy cancels out of the relevant ratio due to the supersymmetry.

\subsection{Multi-string partition function}

The multi-string partition function describing an ideal gas of pp-wave
strings is known to be described in terms of the single string
partition functions of bosonic modes ($Z_1^B$) and spacetime fermionic
modes ($Z_1^F$), 
\begin{equation} 
\ln{Z} (a, b, m) = \sum_{r=1}^\infty \frac{1}{r}
\left( Z_1^B (ar, br, m) - (-1)^r Z_1^F (ar, br, m) \right) \;.
\end{equation}
For the supersymmetric string, the partition functions for the two
modes differ only by a finite number - the number of bosonic minus
fermionic zero modes.  This gives a small, constant
contribution to the free energy, which at high temperatures will be
negligible \cite{pp-wave Hagedorn}.  Thus
\begin{equation}
\ln{Z}(a,b,m) = \sum_{^{r=1}_{r ~odd}}^{\infty}\frac{1}{r}~Z_{1}(ar,br,m).
\end{equation}

Substituting (\ref{single string partition function}) and
(\ref{zlc-theta}) into the above expression, and changing variables of
integration from $p^{+}$ to $\displaystyle \tau_{2}=\frac{ar}{4\pi\alpha'p^{+}}$ now
yields
\begin{equation}
\ln{Z}(a,b,\mu) = \frac{a}{4\pi\alpha'}\int_{-\frac{1}{2}}^{\frac{1}{2}}d\tau_{1}\int_{0}^{\infty}\frac{d\tau_{2}}{(\tau_{2})^{2}}\sum_{^{r=1}_{r ~odd}}^{\infty} \left[\frac{\Theta_{\frac{1}{2},0}(\tau_{1},\tau_{2},\frac{\mu ar}{2\pi\tau_{2}})}{\Theta_{0,0}(\tau_{1},\tau_{2},\frac{\mu ar}{2\pi\tau_{2}})}\right]^{4}e^{-\frac{abr^{2}}{4\pi\alpha'\tau_{2}}},
\end{equation}
which is proportional to the Helmholtz free energy.

\subsection{Hagedorn behaviour}

We now intend to investigate the Hagedorn behaviour of our gas of
pp-wave strings, and so, following Greene {\it et. al.} \cite{pp-wave Hagedorn}, we begin by
searching for an exponential divergence of the density of states.
Towards this end, let us consider the building blocks
$\Theta_{\alpha,\delta}$ in the high energy limit $p^{+}\rightarrow
\infty$ (or $\tau_{2}\rightarrow 0$), in which $\displaystyle \tilde{\mu} =
m\tau_{2} = \tfrac{\mu ar}{2\pi}$ is held fixed. The definition
(\ref{building blocks - def}) yields
\begin{equation}
\ln{\Theta}_{\alpha,\delta}(\tau_{1},\tau_{2},\tfrac{\tilde{\mu}}{\tau_{2}}) = 4\pi\tau_{2} E_{\delta}\left(\tfrac{\tilde{\mu}}{\tau_{2}}\right) +
\sum_{n=-\infty}^{\infty}\ln\left(1 - e^{-2\pi\tau_{2}|\omega_{n+\delta}| + 2\pi i\tau_{1}(n+\delta) + 2\pi i\alpha}\right) + c.c.
\end{equation}
and, defining $x \equiv \tfrac{(n+\delta)\tau_{2}}{\tilde{\mu}}$ and
$\theta \equiv \tfrac{\tau_{1}}{\tau_{2}}$, we see that $\Delta x =
\tfrac{\tau_{2}}{\tilde{\mu}}\Delta n \rightarrow dx$ in the high
energy limit, so that $x$ becomes a continuous variable over which we
can integrate.  Hence, since $\tau_{2}|\omega_{n+\delta}| =
\tilde{\mu}\sqrt{1+x^{2}}$, we obtain
\begin{eqnarray}
\nonumber & \ln\Theta_{\alpha,\delta}(\tau_{1},\tau_{2},\tfrac{\tilde{\mu}}{\tau_{2}}) & \longrightarrow \frac{\tilde{\mu}}{\tau_{2}}\int_{-\infty}^{\infty}dx~
\ln\left(1 - e^{-2\pi\tilde{\mu}\sqrt{1+x^{2}} + 2\pi i\tilde{\mu}\theta x + 2\pi i\alpha}\right) + c.c. \\
&& ~~~~~ \equiv -\frac{\tilde{\mu}}{\sqrt{1+\theta^{2}}\tau_{2}}\left[f(\tilde{\mu},\theta,\alpha) + c.c.\right].
\end{eqnarray}
Expanding out the logarithm, we obtain
\begin{eqnarray}
&& \!\!\!\!\!\!\!\!\!\!\! f(\tilde{\mu},\theta,\alpha) = \sqrt{1+\theta^{2}}\int_{-\infty}^{\infty}dx\sum_{l=1}^{\infty}\frac{1}{l}~e^{-2\pi\tilde{\mu}\sqrt{1+x^{2}} + 2\pi i\tilde{\mu}\theta x 
+ 2\pi il\alpha} + c.c.\nonumber\\
&& ~~~~~~  = 2\sqrt{1+\theta^{2}}\sum_{l=1}^{\infty}\frac{1}{l}~e^{2\pi il\alpha}\int_{0}^{\infty}dx~e^{-2\pi l\tilde{\mu}\sqrt{1+x^{2}}}\cos{(2\pi l\tilde{\mu}\theta x)} + c.c.~~~~~~ \nonumber\\
& & ~~~~~~ =2\sum_{l=1}^{\infty}\frac{1}{l}~e^{2\pi il\alpha}~K_{1}(2\pi\tilde{\mu}l\sqrt{1+\theta^{2}}) + c.c. ~~~~~~~~~~ ~~~~~~~~~~ ~~~~~~~~~~ ~~~~
\label{function-f}
\end{eqnarray}
where $K_{1}(x)$ is a modified Bessel function of the second kind,
which is a real positive monotonically decreasing function tending to
zero quickly as $x\rightarrow \infty$.

Now, since $\bar{f}(\tilde{\mu},\theta,\alpha) =
f(\tilde{\mu},\theta,\alpha)$, when $\alpha = 0$ or $\alpha =
\tfrac{1}{2}$, we find that
\begin{eqnarray}
&& \!\!\!\!\!\! \ln{\Theta_{\frac{1}{2},0}}(\tau_{1},\tau_{2},\tfrac{\tilde{\mu}}{\tau_{2}}) - \ln{\Theta_{0,0}}(\tau_{1},\tau_{2},\tfrac{\tilde{\mu}}{\tau_{2}}) \nonumber\\
\nonumber && \!\!\!\!\!\! \longrightarrow -\frac{2\tilde{\mu}}{\sqrt{1+\theta^{2}}\tau_{2}}\left[f(\tilde{\mu},\theta,\tfrac{1}{2}) - f(\tilde{\mu},\theta,0)\right]
= \frac{8\tilde{\mu}}{\sqrt{1+\theta^{2}}\tau_{2}}\sum_{^{l=1}_{l ~odd}}^{\infty}\frac{1}{l}~K_{1}(2\pi l\tilde{\mu}\sqrt{1+\theta^{2}}).\nonumber
\end{eqnarray}

Thus the behaviour of the multi-string partition function in the
high-energy limit $\tau_{2}\rightarrow 0$ is given by
\begin{equation}
\label{undeformed Z}
\ln{Z}(a,b,\mu) \longrightarrow \frac{a}{4\pi\alpha'}\sum_{^{r=1}_{r ~odd}}^{\infty}\int_{0}^{\infty}\frac{d\tau_{2}}{\tau_{2}}\int_{-\frac{1}{2\tau_{2}}}^{+\frac{1}{2\tau_{2}}}d\theta~ e^{-\frac{abr^{2}}{4\pi\alpha'\tau_{2}} + \frac{16\mu ar}{\pi\tau_{2}}\frac{1}{\sqrt{1+\theta^{2}}}\left[\sum\limits_{^{~l=1}_{~l ~odd}}^{\infty}\frac{1}{l}~K_{1}\left(\mu alr\sqrt{1+\theta^{2}}\right)\right]},\nonumber
\end{equation}
where we have changed the integral over $\tau_{1}$ into an integral
over $\theta = \tfrac{\tau_{1}}{\tau_{2}}$.

We now wish to determine for which temperatures (values of $a$ and
$b$) this partition function converges. Only the $r=1$ term need be
considered\footnote{The modified Bessel function $K_{1}$ is
  monotonically decreasing so that all the $r>1$ terms are much
  smaller (exponentially so) than the $r=1$ term.  Therefore, if the
  $r=1$ term converges, then all the other terms are also
  convergent.}. The convergent/divergent nature of the integral over
$\tau_{2}$ depends critically on the sign of the expression in the
exponential.  The integral converges if
\begin{equation}
ab < \frac{64 a \alpha'\mu}{\sqrt{1+\theta^{2}}}\sum\limits_{^{~l=1}_{~l ~odd}}^{\infty}\frac{1}{l}~K_{1}\left(\mu
al\sqrt{1+\theta^{2}}\right) \leq 64 a \alpha'\mu\sum\limits_{^{~l=1}_{~l
~odd}}^{\infty}\frac{1}{l}~K_{1}\left(\mu al\right) \equiv \beta_{H},
\end{equation}
for all $\theta$. This critical point $ab=\beta_{H}$ corresponds to the Hagedorn temperature $T_{H}$ defined by
\begin{equation}\label{hagedorn temperature}
T_{H}^{-2} = \beta_{H} + a^{2}\mu^{2}\sum_{i=1}^{8}\left(x^{i}\right)^{2} = 64\alpha'\mu
a\sum_{l=1}^{\infty}\frac{1}{l}~K_{1}(\mu al) + a^{2}\mu^{2}\sum_{i=1}^{8}\left(x^{i}\right)^{2}.
\end{equation}

The nature of the partition function at $ab=\beta_{H}$ was considered
in the third of references \cite{pp-wave Hagedorn}, and is related to the behavior of
thermodynamic quantities as they approach the critical point.  A phase
transition requires a finite free energy, though derived quantities,
such as internal energy ($E = -(\ln{Z})'$) or specific heat ($C =
\beta^2 (\ln{Z})''$) may diverge.  The hallmark of a limiting
temperature, on the other hand, is a free energy which blows up near
$\beta_H$. However, even in this case, it has been argued by Atick and Witten \cite{Hagedorn-Refs} that string interactions can turn this into a first
order transition, with a critical temperature below the Hagedorn temperature.
In the high energy limit $\tau_{2}\rightarrow
0$, the integral over $\theta$ is dominated by the saddle point at
$\theta = 0$, with $e^{-c\theta^{2}/\tau^{2}}$ fluctuations ($c$ being
a relatively unimportant constant). Integrating over these
fluctuations will produce a factor of $\sqrt{\tau_{2}}$. Consequently, after performing the integral over
$\tau_{2}$, the multi-string partition function goes like
\begin{equation}
\beta F = -\ln{Z}(a,b,\mu) \propto \sqrt{ab - \beta_{H}} + \textrm{regular},
\end{equation}
which remains finite at $ab=\beta_{H}$, signaling a phase transition.

\section{The deformation}

As discussed in the Introduction, there are two classes of BPS deformations of the maximally symmetric pp-wave background, both arrived at by deforming the $AdS_{5}\times S^{5}$ type IIB solution $-$thereby breaking the $SO(6)$ isometry of the round 5-sphere to $U(1)^{3}$ $-$ and then taking a Penrose limit about an appropriate null geodesic. To summarize\footnote{We refer the reader to \cite{LM} for more details on the deformation and to \cite{Deformed giants} for our notational conventions.  \cite{Mateos} also includes useful derivations.}, the Lunin-Maldacena transformation
maps the $AdS_{5}\times S^{5}$ metric to 
\begin{eqnarray}
  ds^{2} &=& R^{2}{\Bigl[} - \cosh^{2}\rho\,dt^{2} + d\rho^{2} + \sinh^{2}\rho\,d\Omega_{3}^{2} + 
  d\alpha^{2} + G\cos^{2}\alpha\,d\phi_{1}^{2} + \sin^{2}\alpha {\bigl(} d\theta^{2}\nonumber\\
  &+& G\cos^{2}\theta\,d\phi_{2}^{2} + G\sin^{2}\theta\,d\phi_{3}^{2}{\bigr)} + \hat{\gamma}^{2} 
  G\cos^{2}\alpha \sin^{4}\alpha \cos^{2}\theta \sin^{2}\theta \left( d\phi_{1} + d\phi_{2} + d\phi_{3}
  \right)^{2}{\Bigr]},\nonumber
\end{eqnarray}
while turning on the NS $B-$field 
\begin{eqnarray}  
  B_{(2)} &=& \hat{\gamma}R^{2} G {\Bigl(} \sin^{2}\alpha \cos^{2}\alpha \cos^{2}\theta\,d\phi_{1}
  \wedge d\phi_{2} + \sin^{2}\alpha \cos^{2}\alpha \sin^{2}\theta\, d\phi_{3}\wedge d\phi_{1}
  \nonumber\\
  &+& \sin^{4}\alpha \cos^{2}\theta \sin^{2}\theta\, d\phi_{2}\wedge d\phi_{3}{\Bigr)},\nonumber
\end{eqnarray}
and the RR-fluxes
\begin{eqnarray}  
  F_{(3)} &=& -\frac{4\hat{\gamma}}{g_{s}}R^{2}\cos^{2}\alpha \sin^{3}\alpha \cos\theta \sin\theta
  \,d\alpha \wedge d\theta \wedge \left( d\phi_{1} + d\phi_{2} + d\phi_{3}\right),\nonumber\\
  F_{(5)} &=& \frac{4}{g_{5}}R^{4}\left( \cosh\rho \sinh^{3}\rho\, dt\wedge d\rho \wedge d
  \Omega_{3} + G \cos\alpha \sin^{3}\alpha d\phi_{1}\wedge d\alpha \wedge d\widetilde{\Omega}_{3}\right).
\nonumber 
\end{eqnarray}  
Since strings $-$ the objects of interest in this article $-$ couple only to the background geometry and the $B-$field, we can, and will, ignore the RR-sector in what follows. 

\subsection{The pp-wave limit around single-charge null geodesics}

The first of the pp-wave geometries associated to this background arises from taking a Penrose limit about the $(J,0,0)$ null geodesic on $S_{\gamma}^{5}$ by setting
\begin{eqnarray}
 \rho &=& \frac{y}{R},\qquad\qquad\quad\, \alpha = \frac{r}{R},\nonumber\\
 t &=& \mu x^{+} + \frac{x^{-}}{2\mu R},\qquad \phi_{1} = \mu x^{+} - \frac{x^{-}}{2\mu R},\nonumber
\end{eqnarray}
and then scaling $R\rightarrow\infty$. The resulting background fields of the NS sector are
\begin{eqnarray}\label{pp-metric-gamma}
 ds_{\gamma}^{2} &=& -2dx^{+}dx^{-} - \mu^{2}\left[\sum_{i=1}^{4}\left(x^{i}\right)^{2} + (1+\hat{\gamma}^{2})\sum_{i=1}^{4}(y^{i})^{2}\right]\left(dx^{+}\right)^{2}\nonumber\\ 
&+& \sum_{i=1}^{4}\left(dx^{i}\right)^{2} + \sum_{i=1}^{4}\left(dy^{i}\right)^{2},
\end{eqnarray}
and
\begin{eqnarray}
  B_{(2)} &=& \mu \hat{\gamma}r^{2}\left( \cos^{2}\theta\,dx^{+}\wedge d\phi_{2} + \sin^{2}\theta\,d
  \phi_{3}\wedge dx^{+}\right),
\end{eqnarray}
with constant dilaton. The transverse coordinates in the original pp-wave background are
naturally split into two sets of four coordinates (which we shall
denote $x^{i}$ and $y^{i}$) by the self-dual five-form flux. The
effect of the deformation is to alter the effective string mass for
oscillations in \emph{one} set of transverse directions (say $y^{i}$) consequently breaking
the $SO(8)$ degeneracy of the oscillation spectrum. Quantization of the closed string sigma model on this background yields an oscillation spectrum in the eight transverse directions of 
\begin{eqnarray}
&& x^{i}: ~~~ \omega_{n} = \textrm{sign}(n)\sqrt{m^{2} + n^{2}}, \\
&& y^{i}: ~~~ \omega^{\pm}_{n} = \textrm{sign}(n)\sqrt{m^{2} + (n\pm\hat{\gamma} m)^{2}},
\end{eqnarray}
where $m = 2\alpha'\mu p^{+}$ and $\pm$ indicates the spin in the
$y_{1},y_{2}$ or $y_{3},y_{4}$ planes. As expected, fluctuations in the $y^{i}$ directions are dependent on the deformation parameter while those in the $x^{i}$ are not.

\subsection{A homogeneous plane wave limit}
With only a $U(1)^{3}$ isometry remaining on the Lunin-Maldacena background, not all Penrose limits are equivalent as not all geodesics can be rotated into each other. In particular, by focussing on states that live near the null geodesic $\displaystyle \tau = \psi \equiv \frac{1}{3}(\phi_{1}+\phi_{2}+\phi_{3}) $ with $\displaystyle \alpha_{0}=\cos^{-1}(\frac{1}{\sqrt{3}})$ and $\displaystyle \theta_{0}=\frac{\pi}{4}$. Setting\footnote{Here, $\varphi_1 \equiv \frac{1}{3} (\phi_1 + \phi_3 - 2\phi_2)$ and $\varphi_2 \equiv \frac{1}{3} (\phi_1 +\phi_2 - 2\phi_3)$.}
\begin{eqnarray}
 \theta = \frac{\pi}{4} + \sqrt{\frac{2}{3}}\frac{x^{1}}{R}\,,\quad \alpha &=& \alpha_{0} - \frac{x^{2}}{R}\,,\quad \rho = \frac{y}{R}\,,\nonumber\\
 \varphi_{1} = \frac{\tilde{x}^{3}}{R}\,\quad \varphi_{2} = \frac{\tilde{x}^{4}}{R}\,,\quad t &=& \mu x^{+} + \frac{x^{-}}{2 \mu R}\,,\quad \psi = \frac{x^{-}}{2\mu R} - \mu x^{+}\,,\nonumber
\end{eqnarray}
redefining
\begin{eqnarray}
 x^{3} = \sqrt{\frac{2}{3+\hat{\gamma}^{2}}}\left( \tilde{x}^{3} + \frac{1}{2}\tilde{x}^{4}\right)\,,\quad
 x^{4} = \sqrt{\frac{3}{2(3+\hat{\gamma}^{2})}}\tilde{x}^{4}\,,\nonumber
\end{eqnarray}
and taking the $R\rightarrow\infty$ limit gives the pp-wave metric
\begin{eqnarray}
  ds^{2} &=& -2dx^{+}dx^{-} - \mu^2 \left( \sum_{a=5}^{8}(x^{a})^{2} + \frac{4\hat{\gamma}^{2}}{3+\hat{\gamma}^{2}}\left((x^{1})^{2} + (x^{2})^2{}\right)\right)(dx^{+})^{2} + \sum_{a=5}^{8}(dx^{a})^{2}\nonumber\\
  &+& \sum_{i=1}^{4}\,(dx^{i})^{2} + \frac{4\mu \sqrt{3}}{\sqrt{3+ \hat{\gamma}^{2}}}(x^{1}dx^{3}
  +x^{2}dx^{4})dx^{+}\,.
  \label{JJJ-def-pp-wave}
\end{eqnarray}
In this same limit, the remaining fields in the NS sector of the IIB multiplet are
\begin{eqnarray}
  B_{(2)} &=& \frac{\hat{\gamma}}{\sqrt{3}}\,dx^{3}\wedge dx^{4} + \frac{2\mu \hat{\gamma}}{\sqrt{3+\hat{\gamma}^{2}}}\,dx^{+}\wedge\left( x^{1}dx^{4} - x^{2}dx^{3}\right)\,,\nonumber\\
  e^{2\phi} &=& \frac{1}{1+\hat{\gamma}^{2}}e^{2\phi_{0}}\,.\nonumber
 \label{JJJ-multiplet}
\end{eqnarray}
Additionally, this background supports non-vanishing RR $2-$ and $4-$forms which, while they result in rather sophisticated D-brane dynamics on the deformed geometry \cite{Deformed giants}, are irrelevant for our analysis. Closed strings in this background, their supersymmetries and dual gauge theory operators were first studied in \cite{DMSS,Mateos} where it was noticed that a change of coordinates from $x^{-}$ to $x^{-}+\sqrt{3/(3+\hat{\gamma}^{2})}(x^{1}x^{3} + x^{2}x^{4})$ brings the $(J,J,J)$ pp-wave metric into the homogeneous plane wave form \cite{Hpp}
\begin{eqnarray}
 ds^{2} = -2dx^{+}dx^{-} + k_{ij}x^{i}x^{j}\,(dx^{+})^{2} + 2f_{ij}x^{i}dx^{j}dx^{+} + dx^{i}dx^{i}\,,
 \label{Homogeneous}
\end{eqnarray}
where the matrices $\displaystyle k_{ij} = \mu^2 {\rm diag}\left[\frac{4\hat{\gamma}^{2}}{3+\hat{\gamma}^{2}},
\frac{4\hat{\gamma}^{2}}{3+\hat{\gamma}^{2}},0,0,1,1,1,1\right]$ and
\begin{eqnarray}
 f_{ij} = \sqrt{\frac{3 \mu^2 }{3+\hat{\gamma}^{2}}}\left[ 
            \begin{array}{cccccccc}
              0&0&1&0&0&0&0&0\\
              0&0&0&1&0&0&0&0\\
              -1&0&0&0&0&0&0&0\\
              0&-1&0&0&0&0&0&0\\
              0&0&0&0&0&0&0&0\\
              0&0&0&0&0&0&0&0\\
              0&0&0&0&0&0&0&0\\
              0&0&0&0&0&0&0&0
            \end{array} \right]\,.
\end{eqnarray}
Remarkably, even though the background and associated string equations of motion depend on $\hat{\gamma}$ in a fairly non-trivial way, the quantum closed string spectrum, 
\begin{eqnarray}
  \omega_{n} = 1 \pm \sqrt{1 + 4n^{2}}\,,
  \label{JJJ-spectrum}
\end{eqnarray} 
determined by the frequency base ansatz of Blau {\it et.al.} \cite{Hpp} for strings on homogeneous plane waves, exhibits no dependence on the deformation \cite{DMSS, Mateos} at all. Consequently, we expect that the high temperature behaviour of an ensemble of strings on this particular deformation of the maximally symmetric pp-wave should be identical to that of homogeneous plane wave strings (see for example \cite{Bigazzi}). While the Hagedorn behaviour of strings on this particular class of homogeneous plane waves ({\it i.e.} non-trivial $k_{ij}$ and $f_{ij}$) has not yet - to the best of our knowledge - been studied, it is clear that it will be independent of the deformation.

\subsection{$\gamma$-deformed $(J,0,0)$ Hagedorn temperature}

Returning again to the $(J,0,0)$ Penrose limit of $AdS_{5}\times S^{5}_{\gamma}$, we shall now 
construct the partition function for an ideal gas
of strings in this $\gamma$-deformed background. 
In computing the partition function, much of the analysis is identical to the undeformed case.  The difference comes from the modified string spectra - there are four real oscillators 
with $|\omega_n| = \sqrt{m^2 + n^2}$, and two each with $| \omega_n^\pm | = \sqrt{m^2 + (n \pm \hat{\gamma} m)^2}$.
This does, in fact, lead to a partition function with non-trivial $\hat{\gamma}-$dependence. 
Remarkably though, we will see that in the high temperature limit ($\tau_2 \to 0$), this
difference disappears, and the actual Hagedorn temperature itself is undeformed.

At the level of the building blocks, the effect of the $\gamma$-deformation is to change two of 
the $\Theta_{\alpha,\delta}$ to
\begin{eqnarray}\label{deform building blocks}
\nonumber & \Theta_{\alpha,\delta}^{\pm}(\tau_{1},\tau_{2},m) & \equiv e^{4\pi\tau_{2}E^\pm_{\delta}(m)}\prod_{n=-\infty}^{\infty}
\left(1 - e^{-2\pi\tau_{2}|\omega^\pm_{n+\delta}| + 2\pi i\tau_{1}(n+\delta) + 2\pi i \alpha}\right) \\
&& ~~~~~~~~~~ ~~~~~~~~~~ ~~ \times \left(1 - e^{-2\pi\tau_{2}|\omega^\pm_{n-\delta}| + 2\pi i\tau_{1}(n-\delta) - 2\pi i \alpha}\right).~~~~~~
\end{eqnarray}
The exact form of the energy $E^{\pm}_{\delta}(m)$ is unimportant
in our present discussion (as long as it is still independent of
$\alpha$), since it cancels out of the relevant ratio of building
blocks due to the residual supersymmetry.

The $\gamma$-deformed $(J,0,0)$ multi-string partition function can
now be written (in the same fashion as (\ref{undeformed Z})) as
\begin{eqnarray}
\ln{Z}^{\gamma}(a,b,\mu) & = &
\frac{a}{4\pi\alpha'}\int_{-\frac{1}{2}}^{\frac{1}{2}}d\tau_{1}\int_{0}^{\infty}\frac{d\tau_{2}}{(\tau_{2})^{2}} \nonumber \\
& \times &
\sum_{^{r=1}_{r ~odd}}^{\infty} e^{-\frac{abr^{2}}{4\pi\alpha'\tau_{2}}} 
\left( \frac{\Theta_{\frac{1}{2},0}}{\Theta_{0,0}}\right)^{2}
\left( \frac{\Theta^{+}_{\frac{1}{2},0}}{\Theta^{+}_{0,0}} \right)
\left( \frac{\Theta^{-}_{\frac{1}{2},0}}{\Theta^{-}_{0,0}}\right) \;,
\end{eqnarray}
where each $\Theta$ is an implicit function
$\displaystyle \Theta_{\alpha,\delta}\left(\tau_{1},\tau_{2},\frac{\mu ar}{2\pi\tau_{2}}\right)$.

Following the steps in section 2, it is not hard to show that in the high-energy limit, 
the deformed oscillators lead to 
the replacement of the function $f$ in
(\ref{function-f}) with 
\bea
  f_\gamma^\pm = 2 
  \sum_{l=1}^\infty \frac{e^{2\pi i l (\alpha \mp \hat{\gamma} \tilde{\mu} \theta)}}{l} K_1 (2  
  \pi \tilde{\mu} l \sqrt{1+\theta^2}) 
  \equiv  f(2\pi \tilde{\mu} l \sqrt{1+\theta^2}, \alpha \mp
  \hat{\gamma} \tilde{\mu} \theta) \;.  
\eea 
In contrast to the undeformed case, here we cannot set $\bar{f} = f$.  Instead, for
$\alpha = 0,1/2$,
\bea
  \bar{f}(x, \alpha \mp \hat{\gamma} \tilde{\mu} \theta) = f (x,
  \alpha \pm \hat{\gamma} \tilde{\mu} \theta) \;.
\eea
Consequently, 
\begin{eqnarray}
\nonumber & \!\!\!\! \left[\ln{\Theta}^{\gamma}_{\frac{1}{2},0} -
\ln{\Theta}^{\gamma}_{0,0}\right](\tau_{1},\tau_{2},\tfrac{\tilde{\mu}}{\tau_{2}}) &\longrightarrow
-\frac{2\tilde{\mu}}{\sqrt{1+\theta^{2}}\tau_{2}}\left[\tfrac{1}{2}f(\tilde{\mu},\theta,\tfrac{1}{2}) +
\tfrac{1}{4}f^{\gamma}_{+}(\tilde{\mu},\theta,\tfrac{1}{2}) +
\tfrac{1}{4}f^{\gamma}_{-}(\tilde{\mu},\theta,\tfrac{1}{2}) \right. \\
\nonumber && \left. ~~~~~~~~~~ ~~~~~~~~~~ - \tfrac{1}{2}f(\tilde{\mu},\theta,0) -
\tfrac{1}{4}f^{\gamma}_{+}(\tilde{\mu},\theta,0) - \tfrac{1}{4}f^{\gamma}_{-}(\tilde{\mu},\theta,0)\right]\\
\nonumber && = -\frac{4\tilde{\mu}}{\sqrt{1+\theta^{2}}\tau_{2}}\sum_{^{l=1}_{l ~odd}}^{\infty}\frac{1}{l}\left[1 +
\cos{(2\pi l\hat{\gamma}\tilde{\mu}\theta)}\right]K_{1}(2\pi l\tilde{\mu}\sqrt{1+\theta^{2}}),\nonumber
\end{eqnarray}

and the diverging piece of the partition function (dominated by high energy modes) becomes
\begin{eqnarray}
 && \!\!\!\! \ln{Z}^{\gamma}(a,b,\mu) \nonumber\\
\nonumber && \!\!\!\! \longrightarrow \frac{a}{4\pi\alpha'}\sum_{^{r=1}_{r
~odd}}^{\infty}\int_{0}^{\infty}\frac{d\tau_{2}}{\tau_{2}}\int_{-\frac{1}{2\tau_{2}}}^{\frac{1}{2\tau_{2}}}d\theta~
e^{-\frac{abr^{2}}{4\pi\alpha'\tau_{2}} + \frac{8\mu
ar}{\pi\tau_{2}}\frac{1}{\sqrt{1+\theta^{2}}}\left[\sum\limits_{^{~l=1}_{~l
~odd}}^{\infty}\frac{1}{l}\left[1+\cos{(\mu arl\theta\hat{\gamma})}\right] K_{1}\left(\mu
alr\sqrt{1+\theta^{2}}\right)\right]}.
\end{eqnarray}

Despite these changes to the partition function, in evaluating the
high energy behavior, the $\tau_1$ (i.e., $\theta)$ integral is dominated by
a Gaussian which picks out $\theta=0$; 
in this limit, 
the evaluation of the free energy is identical to the undeformed case\footnote{In evaluating the gaussian, only the width changes - this affects the proportionality constant for $F$, but not the location of the singularity.}
\be 
  -\beta F \sim \sqrt{\beta^2 - \beta_H^2} +
  \mathrm{regular} \;, 
\ee with $\beta^2 = ab$ and 
\be 
    \beta_H^2 =  64\alpha'\mu\sum\limits_{^{~l=1}_{~l
~odd}}^{\infty}\frac{1}{l}~K_{1}\left(\mu al\right) \;.  
\ee
so that the Hagedorn temperature once more describes a phase
transition. To summarize, a key feature of this computation is that in the high temperature limit we find a continuum of
states where $x\equiv(n+\delta)/\tilde{\mu}$ is effectively continuous. The spacetime deformation manifests in the partition function only in that this continuous variable is changed from $x \to x \mp \hat{\gamma}$ and since the Hagadorn temperature is given by the density of states $\rho(w) = \left(dw(n)/dn\right)^{-1}$, it must remain unchanged even though the spectrum of strings  - and, consequently the partition function also - on this background depends rather non-trivially on $\hat{\gamma}$.

\section{Matching the deformed gauge and string theories.} 
A direct comparison between the thermodynamic properties of pp-wave strings
(deformed or otherwise) and the corresponding SYM operators is quite
non-trivial, largely because the pp-wave background is constructed by taking a
Penrose limit in which the radius $R$, and hence also the t'Hooft
coupling $\lambda \sim R^{4}$, is large. 
More precisely, the correspondence identifies the momenta $p^{\pm}$
of pp-wave string states with the conformal dimensions, $\Delta$, and 
$U(1)$ R-charges, $J$, of the SYM operators via
\begin{eqnarray}
  \frac{2p^{+}}{\mu} = \Delta - J,\qquad 2\mu\alpha'p^{-} = \frac{\Delta + J}{\sqrt{\lambda}}\,,
\end{eqnarray}  
so that in the $N\to\infty$ $\lambda\to\infty$ limit with $p^{\pm}$ finite, the only states that survive are those with conformal dimension and R-charge that scale like $\sqrt{N}$. Since these are precisely the gauge theory states conjectured to be dual to the pp-wave string \cite{BMN}, the problem with matching the Hagedorn/deconfinement temperature of the guage theory to the Hagedorn temperature of the string theory is evident: BMN states form only a small subset of the set of all possible states in the gauge theory. At small 't Hooft coupling all of these states are manifest in the gauge theory resulting in an apparent gross mismatch with a state counting on the string side where the number of states grows exponentially as the Hagedorn temperature is approached. 
 
\subsection{A novel decoupling limit}
Armed with the latest in AdS/CFT technology, it was suggested in the series of works \cite{Matching Hagedorns} that this problem may be (at least partially) resolved by a new decoupling limit of the correspondence.
In the gauge theory, this decoupling corresponds to low temperatures and near-critical 
chemical potentials while decoupling in the string theory
takes place in the large $\mu$ limit of a particular compactified pp-wave background with a flat direction.
In the interests of self-containment, we summarize here the main results in this argument.

The $SU(N)$, $\mathcal{N}=4$ Yang-Mills supermultiplet consists of 
three complex scalars $X,Y$ and $Z$, four Weyl spinors
$\psi^{\alpha}_{i}$ and one gauge vector boson $A_{\mu}$. The
partition function is a sum over all multi-trace operators 
\begin{eqnarray}
 {\cal O} = {\rm Tr}\left(W_{1}^{(1)}\ldots W_{l_{1}}^{(1)}\right)\ldots {\rm Tr}\left(W_{1}^{(k)}\ldots W_{l_{k}}^{(k)}\right),
 \label{multi-trace}
\end{eqnarray}
constructed from these fields or their derivatives. In the Harmark-Orselli decoupling limit\footnote{See \cite{Matching Hagedorns} for a more detailed discussion.}
which sends $T, \lambda, \epsilon \to 0$ while keeping $\tilde{T} = T/\epsilon$ and $\tilde{\lambda} = \lambda/\epsilon$ fixed,
most of the ${\cal N}=4$ SYM states decouple - only those whose bare dimension is equal to their $R$-charge survive.  This reduces to a thermal quantum mechanics system with partition function
\begin{equation}
Z(\tilde{\beta}) = \textrm{Tr}\left(e^{-\tilde{\beta}( D_{0} + \tilde{\lambda}D_{2})}\right) 
\end{equation}
- $D_0$ and $D_2$ are the tree and 1-loop contributions of the dilatation operator. 
This leaves behind the well-known $SU(2)$ sector, where only the two complex Higges, $Z$ and $X$ contribute to the multi-trace operator (\ref{multi-trace}).
In principle then, since $D_{2}$ is known, $Z(\tilde{\beta})$ is computable for any value of $\lambda$ and $N$. In practice, it is easier to take the planar limit $N\rightarrow \infty$, in which single trace operators dominate and $D_{2}$ maps to the Hamiltonian of the spin-$\frac{1}{2}$ XXX-Heisenberg spin chain.  The partition function for a thermal state of the Yang-Mills theory can then be computed as
\begin{eqnarray}
  Z(\tilde{\beta}) = \exp\left[\sum_{n>0}\sum_{l>1}\frac{1}{n}e^{-n\tilde{\beta}l}Z_{l}^{(XXX)}(n\tilde{\beta})\right],
\end{eqnarray}
where, following \cite{Matching Hagedorns}, $Z_{l}^{(XXX)}(n\tilde{\beta})$ denotes the partition function of the ferromagnetic spin chain of length $l$.

Key to this matching prescription is the manifest spatial isometry in the dual string background about which the decoupling limit is taken; a pp-wave geometry with a flat direction. In our notation, with two of the 5-sphere angular momenta turned on, this is precisely the $(J,J,0)$ pp-wave,
\begin{eqnarray}
  ds^{2} = -2dx^{+}dx^{-} + \sum_{i=1}^{8}\left(dx^{i}\right)^{2} - 
  \mu^{2}\sum_{i=3}^{8}\left(x^{i}\right)^{2}\left(dx^{+}\right)^{2} - 4\mu x^{2}dx^{1}dx^{+}\,.
\end{eqnarray}
Although this background is related to the $(J,0,0)$ one by a time dependent coordinate rotation 
\begin{eqnarray}
   \left[\begin{array}{c}
             y^{1}\\
             y^{2}
          \end{array}\right] = 
   \left[\begin{array}{cc}
             \cos(\mu x^{+}) & -\sin(\mu x^{+})\\
             \sin(\mu x^{+}) & \cos(\mu x^{+})
          \end{array}\right]
   \left[\begin{array}{c}
             x^{1}\\
             x^{2}
          \end{array}\right]\nonumber             
\end{eqnarray}
in the $(x^{1},x^{2})-$plane, the physics is rather different. In particular, there is  one vacuum state for each value of the momenta along the flat direction, $x^{1}$ so that in the limit where $\epsilon\to0$ with $\tilde{\mu} = \mu\sqrt{\epsilon}$, $\tilde{H}_{\rm lc} = H_{\rm lc}/\epsilon$, $\tilde{g}_{s} = g_{s}/\epsilon$,  $l_{\rm s}$ and $p^{+}$ all fixed, the pp-wave spectrum - and consequently the Hagedorn behaviour also - exactly matches the weakly coupled gauge theory.  At this point, everything we have said so far applies specifically to the maximally supersymmetric ${\cal N}=4$ SYM theory. How then, is this matching prescription affected under a systematic deformation, such as the Leigh-Strassler deformation, away from maximal supersymmetry? 

Very generally, the Leigh-Strassler deformation \cite{Leigh-Strassler} of ${\cal N}=4$ SYM produces a three-parameter family of field theories that all preserve ${\cal N}=1$ supersymmetry with the ${\cal N}=4$ superpotential mapping to 
\begin{eqnarray}
  h\,{\rm tr}\left(e^{i\pi\beta}\Phi_{1}\Phi_{2}\Phi_{3} - e^{-i\pi\bar{\beta}}\Phi_{1}\Phi_{3}\Phi_{2}\right)
  + h'\,{\rm tr}\left(\Phi_{1}^{2}+\Phi_{2}^{2}+\Phi_{3}^{2}\right)
\end{eqnarray}
Within this class of theories, of particular interest to us is the case $(h,h') = (1,0)$ and $\beta = \bar{\beta}\equiv \gamma$. For this choice, the Leigh-Strassler superpotential deformation can be resummed as a Moyal-like $*-$product deformation $\displaystyle \Phi_{1}*\Phi_{2} = e^{i\pi\gamma(Q^{1}_{\Phi_{1}}Q^{2}_{\Phi_{2}} - Q^{2}_{\Phi_{1}}Q^{1}_{\Phi_{1}})}\Phi_{1}\Phi_{2}$, where $(Q^{1}_{\Phi_{i}},Q^{2}_{\Phi_{i}})$
are the charges of the $\Phi_{i}$ fields under a global $U(1)_{1}\times U(1)_{2}$ symmetry of the Yang-Mills theory \cite{LM}. Consequently, not only is the Feynman diagram structure (at the planar level) unchanged from the undeformed theory but since the deformation also preserves the three Cartan generators of the $SO(6)$ R-symmetry of the Yang-Mills theory, any closed subset of single-trace operators in the original theory remains closed under the renormalization group flow in the Lunin-Maldacena deformation. Specifically, this is true of the $SU(2)$ and $SU(3)$ sectors\footnote{Since the $U(1)$ sector of the theory is spanned by single-trace operators constructed from just one of the complexified Yang-Mills scalars, it is a straightforward consequence of the holomophicity of these operators that this sector remains unaffected by the deformation} consisting of single-trace operators built out of two and three complex Higgs fields in the SYM supermultiplet respectively. 

Indeed, like its undeformed counterpart, the dilatation operator of this $\gamma-$deformed field theory can also be represented as a Hamiltonian of a spin-chain acting on a spin-chain Hilbert space\footnote{For the sake of definiteness and to facilitate comparison with the (undeformed) Hagedorn/phase-transition analysis of \cite{Matching Hagedorns}, we now restrict our attention to the $SU(2)_{\gamma}$ sector of the $\mathcal{N}=1$ theory and content ourselves with some comments on the $U(1)$ and $SU(3)$ sectors at the end of this section.}. 

To reiterate, under the Lunin-Maldacena deformation, since all commutators
$[A,B]\rightarrow [A,B]_{\gamma} \equiv e^{i\pi\gamma}AB - e^{-i\pi\gamma}BA$, interchanging any two differently charged letters in a single-trace operator ${\rm Tr}(X^{J_{1}}Y^{J_{2}})$ comes with a $\gamma-$dependent phase. At the level of the associated spin-chain Hamiltonian, this deformation can be realized \cite{deformed-spin-chain} either a parity-preserving ferromagnetic XXZ spin-chain with $\gamma-$twisted boundary conditions or as the Hamiltonian of a XXZ spin-chain with broken parity and periodic boundary conditions,
\begin{eqnarray}
  H &=& \frac{\lambda}{(4\pi)^{2}}\sum_{l=1}^{J}
  \left[\mathbb{I}_{l}\otimes\mathbb{I}_{l+1} - \left(\sigma^{x}_{l}\otimes \sigma^{x}_{l+1} + \sigma^{y}_{l}\otimes \sigma^{y}_{l+1} + \sigma^{z}_{l}\otimes \sigma^{z}_{l+1}\right)\right.\nonumber\\
  &{}&\\
   &+& \left.\left(1 - \cos(2\pi\gamma)\right)\left(\sigma^{x}_{l}\otimes \sigma^{x}_{l+1} + \sigma^{y}_{l}\otimes \sigma^{y}_{l+1}\right) + \sin(2\pi\gamma)\left(\sigma^{x}_{l}\otimes \sigma^{y}_{l+1} - \sigma^{y}_{l}\otimes \sigma^{x}_{l+1}\right)\right]\,.\nonumber
\end{eqnarray}
Either way, the resulting spin-chain lends itself to a Bethe ansatz-type solution (see, for instance, the second of \cite{deformed-spin-chain}) from which the corresponding energy spectrum may be extracted and, following \cite{Matching Hagedorns}, the Hagedorn temperature determined. In principle then, we should be able to match the temperature of the Hagedorn transition of the gauge theory to the Hagedorn temperature of the dual string theory. Or should we? 

The problem is that the $SU(2)$ sector of the gauge theory - the first non-trivial sector in which the matching prescription works - corresponds, in our notation, to a deformed $(J,J,0)$ pp-wave. In the undeformed case, this geometry obtains from $AdS_{5}\times S^{5}$ by expanding around the $(J,J,0)$ null geodesic
\begin{eqnarray}
  \phi^{+} = t,\quad \alpha = \frac{\pi}{2},\quad \theta = \frac{\pi}{4},\quad 
  \phi^{-} = \rho = 0.
\end{eqnarray}
Indeed, setting 
\begin{eqnarray}
  t&\equiv& \mu x^{+} + \frac{x^{-}}{\mu R^{2}},\quad \rho \equiv \frac{r}{R},
  \quad \phi^{-}\equiv\frac{x^{1}}{R},\nonumber\\
  \phi^{+}&\equiv& \mu x^{+} - \frac{x^{-}}{\mu R^{2}},\quad \alpha\equiv 
  \frac{\pi}{2} + \frac{\tilde{r}}{R}\,\quad \theta\equiv\frac{\pi}{4} + 
  \frac{x^{2}}{R},\nonumber
\end{eqnarray}
and scaling $R\rightarrow\infty$ produces the $(J,J,0)$ pp-wave
\begin{eqnarray}
  ds^{2} = -2dx^{+}dx^{-} + \sum_{i=1}^{8}\left(dx^{i}\right)^{2}
  -\mu^{2}\sum_{i=3}^{8}\left(x^{i}\right)^{2}\left(dx^{+}\right)^{2}
  - 4\mu x^{2}dx^{1}dx^{+},
\end{eqnarray}
with an explicit isometry along the flat $x^{1}$ direction. Like the $(J,J,J)$ pp-wave, this background is related to the maximally supersymmetric one by a time-dependent rotation - this time in the $(x^{1},x^{2})-$plane. 
Under the Lunin-Maldacena deformation (or equivalently, the TsT transformation), this metric maps to the deformed pp-wave (we omit several variable re-definitions for brevity)
\begin{eqnarray}
  ds^{2} &=& -2dx^{+}dx^{-} + \sum_{i=1}^{8}\left(dx^{i}\right)^{2}
  \nonumber\\
  &-& \mu^{2}\left[\sum_{i=5}^{8}\left(x^{i}\right)^{2}
  + \left(\frac{4-\hat{\gamma}^{2} - 
  \hat{\gamma}^{4}}{4+\hat{\gamma}^{2}}\right)\sum_{i=3}^{4}
  \left(x^{i}\right)^{2} - \frac{4\hat{\gamma}^{2}}{4
  +\hat{\gamma}^{2}}\left(x^{2}\right)^{2}\right]\left(dx^{+}\right)^{2}
  \qquad\\
  &+& 2\mu\left(x^{1}dx^{2} - x^{2}dx^{1}\right)dx^{+} + 
  \frac{2\hat{\gamma}^{2}}{\sqrt{4+\hat{\gamma}^{2}}}\mu
  \left(x^{3}dx^{4} - x^{4}dx^{3}\right)\,.\nonumber
\end{eqnarray}
Like the $(J,J,J)$ background, this pp-wave is also rotationally disconnected from the deformed $(J,0,0)$ pp-wave. In fact, the situation here is slightly worse; although the Penrose limit is certainly well-defined at the level of the metric, this background is actually non-BPS. The first manifestation of this fact arises when we try to apply the Penrose limit to the NS $B-$field. Beginning with 
\begin{eqnarray}
  B &=& \hat{\gamma}R^{2}G \Bigl[\cos^{2}\alpha\sin^{2}\alpha\cos(2\theta)\,
  d\phi_{1}\wedge d\phi_{+}\nonumber\\
  &+&\cos^{2}\alpha\sin^{2}\alpha\,
  d\phi_{1}\wedge d\phi_{-}
  +  \frac{1}{2}\sin^{4}\alpha\sin^{2}(2\theta)\,d\phi_{-}\wedge d\phi_{+}\,\Bigr],
\end{eqnarray}  
where $\displaystyle \phi_{\pm} \equiv \frac{1}{2}\left(\phi_{2}\pm\phi_{3}\right)$, expanding around the $(J,J,0)$ null geodesic and organizing the series in $R$ we find that to leading order $\displaystyle B = \frac{1}{2}\hat{\gamma}\mu R\,dx^{1}\wedge dx^{+}$ which diverges linearly with $R$ in the Penrose limit. The consequences are clear; if any comparison with the $SU(2)$ sector of the Leight-Strassler deformation of $\mathcal{N}=4$ SYM is to be made, another way must be found than the direct computation of the Hagedorn temperature of strings propagating in the deformed $(J,J,0)$ pp-wave. To date, we have not managed to do so but, given the success of the program advocated in \cite{Matching Hagedorns}, it would be disappointing indeed if this were not possible for the $\mathcal{N}=1$ theory!
\subsection{Decoupling the 1/2 - BPS sector of ${\cal N}=1$ SYM}
To conclude this section, we make a few brief comments about the $U(1)$-sector of the gauge theory. Under the Leigh-Strassler deformation, the gauge theory partition function becomes
\begin{equation}
Z(\beta,\Omega_{i}) = \textrm{Tr}\left(e^{-\beta D^{\gamma} + \beta\sum\limits_{i=1}^{3}R_{i}\Omega_{i}}\right).
\end{equation}
Here $D^{\gamma}$ is the $\gamma$-deformed dilatation operator, while
$\Omega_{i}$ are the three chemical potentials associated with the
$R$-charges, $R_{i}$. At weak coupling $\lambda \ll 1$, the dilatation
operator takes the form $D^{\gamma} = D_{0} + \lambda D_{2}^{\gamma}$
to leading order in $\lambda$.  The $\gamma$ deformation affects only
the interactions, so $D_{0}$ yields simply the bare scaling dimensions.
The $(J,0,0)$ decoupling limit, in which we are particularly interested here, corresponds to the choice of chemical potentials 
$(\Omega_{1},\Omega_{2}, \Omega_{3}) = (\Omega,0,0)$, with $\Omega
\rightarrow 1$.  We hold fixed $\tilde{\beta}\equiv(1-\Omega)\beta$
and $\displaystyle \tilde{\lambda}=\tfrac{\lambda}{1-\Omega}$, resulting in small
temperature/coupling. With this, the partition function becomes
\begin{equation}
Z(\tilde{\beta}) = \textrm{Tr}\left(e^{-\tilde{\beta}(D_{0}+\tilde{\lambda}D_{2}^{\gamma})}\right),
\end{equation}
where the trace now runs over only those multi-trace operators with
$D_{0} = R_{1}$.

Now the only surviving states in the Hilbert space are built out of a
Fock space of single trace operators of the form
$\textrm{Tr}(\Phi_{1}*\Phi_{1}*\ldots*\Phi_{1}) = \textrm{Tr}(\Phi_{1}^{L})$. Clearly holomorphic, these single trace, half-BPS
operators are protected by supersymmetry and therefore vanish under
the action of $D_{2}^{\gamma}$ (as well as all higher order terms in
the $\gamma$-deformed dilatation operator). Hence, confining ourselves
to single-trace operators, the $\gamma$-deformed partition function in
the $(J,0,0)$ decoupling limit is given by
\begin{equation}\label{partition-gauge}
Z_{1}(\tilde{\beta}) = \textrm{Tr}\left(e^{-\tilde{\beta}L}\right) = \sum_{L=1}^{\infty}e^{-\tilde{\beta}L} = \frac{1}{1-e^{-\tilde{\beta}}},
\end{equation}
which is obviously independent of the deformation parameter $\gamma$.

\subsection{$\gamma$-deformed string theory in the $(J,0,0)$ decoupling limit}
 To complete the story, we still need to show that the matching prescription of \cite{Matching Hagedorns} goes through for the deformed string theory as well. Towards this end, on the string theory side the Harmark-Orselli decoupling limit is implemented by taking $\epsilon \rightarrow 0$ (to be identified
with the small parameter $1-\Omega$ in the gauge theory) while keeping
$\tilde{R}^{4} \equiv \tfrac{R^{4}}{\epsilon}$ fixed. A modified
Penrose limit, in which $\tilde{R}\rightarrow \infty$, but $R^{4} \sim
\lambda$ still remains small, may then be imposed, while expanding
around the $(J,0,0)$ null geodesic. The resulting $\gamma$-deformed
pp-wave metric is identical to the unmodified version
(\ref{pp-metric-gamma}), except for an overall factor of
$\sqrt{\epsilon}$. Furthermore, when taking this decoupling limit
$\epsilon \rightarrow 0$, we hold fixed
\begin{equation}
\tilde{\mu}\equiv\mu\sqrt{\epsilon}, ~~~~~~ \tilde{H}_{lc}\equiv\frac{H_{lc}}{\epsilon},  ~~~~~~
\alpha' ~~~ \textrm{and} ~~~ p^{+},
\end{equation}
so that the mass parameter $\mu$ becomes large.

Now, from the $\gamma$-deformed modified pp-wave metric, we can deduce
that the rescaled spectrum for strings polarized in
the eight transverse directions is
\begin{eqnarray}
&& x^{i}: ~~~ \omega_{n} = \frac{1}{\epsilon}~\textrm{sign}(n)\sqrt{\tilde{m}^{2} + \epsilon n^{2}}, \\
&& y^{i}: ~~~ \omega^{\pm}_{n} = \frac{1}{\epsilon}~\textrm{sign}(n)\sqrt{\tilde{m}^{2} + (\sqrt{\epsilon}n\pm\hat{\gamma} \tilde{m})^{2}},
\end{eqnarray}
where $\tilde{m}=m\sqrt{\epsilon}$ is fixed.  Notice that all these
modes go like $\tfrac{1}{\epsilon}$ as $\epsilon\rightarrow 0$. Thus,
using an argument similar to that of \cite{Matching Hagedorns}, we conclude that it
is not possible to excite any of the transverse modes in the
$\epsilon\rightarrow 0$ limit as they correspond to states of infinite
energy. Therefore, in the decoupling limit, no non-trivial excited
modes survive.

The $\gamma$-deformed single string partition function in this $(J,0,0)$ decoupling limit must hence be simply a function of the light-cone momentum $p^{+}$ as follows:
\begin{equation}\label{partition-string}
Z_{1}(b) = \int_{0}^{\infty}dp^{+}~e^{-bp^{+}} = \frac{1}{b}.
\end{equation}
What follows, then, is the following expression for the variable $b$ as a function of the inverse temperature $\beta$:
\begin{equation}
b = 1 - e^{-\tilde{\beta}}, ~~~~~~~~ \textrm{with} ~~~ \tilde{\beta} \equiv \epsilon\beta.
\end{equation}

\section{Conclusion}
The idea that there exist quantities on both sides of the gauge theory/gravity correspondence that are manifestly independent of the amount of supersymmetry or conformal symmetry is certainly not new. That one of these quantities might be the Hagedorn temperature of strings on a particular background is intriguing. As far as we are aware, the first such study of the universality of the Hagedorn behavior of strings on pp-wave geometries was carried out in \cite{Gursoy}. There it was demonstrated that the Hagedorn temperature, $T_{H}$, of strings in the Lunin-Maldacena deformation of a pp-wave limit of the Maldacena-Nu$\tilde{\rm n}$ez solution \cite{MN} of type IIB string theory is independent of $\gamma$. This is, however, only a {\it necessary} condition on the universality of $T_{H}$ that needs to be supplemented by additional arguments before it can be labelled universal. 

In this article, we have pursued and confirmed - at least on the gravity side - the line of reasoning initiated in \cite{Gursoy}; the Hagedorn behavior of strings on different pp-wave backgrounds related by an integrable supersymmetric deformation is exactly the same. This lends further support to the conjecture that on the field theory side, in the large R-charge limit, the Hagedorn behavior of  $\mathcal{N}=4$ SYM and its $\mathcal{N}=1$ Leigh-Strassler deformation, two different field theories, is the same as well. The physics of this $\gamma-$independence is essentially the same as that reported in \cite{Gursoy}; in the UV the string excitation number, from which the Hagedorn temperature is determined, is effectively a continuous variable and the only effect of the deformation is to shift $n$ by $\gamma$. On the gauge theory side, utilizing technology developed in the series of papers \cite{Matching Hagedorns}, we have explored the possibility that the matching of Hagedorn and Confinement/Deconfinement via the Heisenberg spin chain might be extended to the $\mathcal{N}=1$ theory. Unfortunately, our results indicate that this may be significantly more subtle than the $\mathcal{N}=4$ case. As we have argued in the last two sections, the $U(1)$ sector, spanned by holomorphic $\frac{1}{2}-$BPS operators, is unchanged by the deformation. Consequently, we have shown that the matching of Hagedorn behavior goes through unaffected. The $SU(2)$ sector, on the other hand, is far from trivial. Under the Leigh-Strassler deformation, the XXX Heisenberg spin chain associated to single-trace operators in this sector is mapped to an XXZ spin chain whose Hamiltonian may be diagonalized by an appropriate Bethe ansatz. Even though the deconfinement transition temperature may then be computed, we argue that no matching with the string theory may be made, along the lines of \cite{Matching Hagedorns} since the corresponding dual geometry is ill-defined. 

It is clear then that this study of the thermal properties of strings on these deformed pp-wave backgrounds and their dual gauge theories generates many more questions than answers. Among those that we think deserve greater future attention are
\begin{itemize}
 \item 
  A more detailed study of the Hagedorn behavior of strings on the $(J,J,J)$
  homogeneous plane wave: Certainly, as we have argued in the main text, we expect
  that the Hagedorn temperature will be independent of the deformation 
  parameter. Nevertheless, it remains to be seen what the actual 
  Hagedorn behavior of strings on the homogeneous plane wave is. 
  Once determined, this temperature
  may be compared to the 
  confinement/deconfinement transition temperature in the $PSU(2|3)$ sector 
  of the deformed gauge theory \cite{deformed-spin-chain} following 
  the prescription of \cite{Matching Hagedorns}.
 \item
  The Hagedorn behavior of strings under deformations of $AdS_{5}$: 
  While the focus of this article was primarily
  on supersymmetric deformations of $AdS_{5}\times S^{5}$ it should be 
  noted that the set of all
  integrable deformations is actually quite large. In addition to complex 
  $\beta-$deformations and the nonsupersymmetric $3-$parameter deformation
  that results in the Frolov-Roiban-Tseytlin background of \cite{FRT}, the
  TsT transformation of \cite{Frolov} has also been applied to the global 
  toroidal isometries of the $AdS_{5}$ part of the 10-dimensional geometry
  \cite{Swanson}. Even though strings propagating on this 
  $AdS_{5}^{\gamma}\times S^{5}$ geometry retain many of the properties 
  of the Lunin-Maldacena background including - but not restricted to - classical
  integrability and a sensible pp-wave limit, the dual field theory seems to be more
  complicated and resembles a fully non-commutative Yang-Mills theory. 
  Nevertheless, the thermodynamic properties remain well defined and it would be  
  of interest to examine the Hagedorn behavior of strings on this background
  as well as its pp-wave limit for signs of universality.
 \item
  Finally, it would be of obvious interest to know whether the Hagedorn behavior
  reported here (and in \cite{Gursoy}) persists for other geometries than
  $AdS_{5}\times S^{5}$. The Klebanov-Strassler background \cite{KS} 
  for instance, would make an excellent example, not only for its confining 
  properties but also because its $\gamma-$deformation is easily constructed 
  \cite{LM}.  
   
\end{itemize}

Barring some fundamentally new insight into the nature of gravity or gauge theories at strong coupling, the AdS/CFT conjecture will likely remain just that, a conjecture, at least for the foreseeable future. In the absence of such a breakthrough though, concepts like integrability and universality are becoming an increasing part of the gauge/gravity vocabulary. However, whereas an enormous amount of energy has in recent years been devoted to unveiling the integrability structures on both sides of the duality (with some truly remarkable successes); the search for universal properties of the correspondence seems to have been less systematic. It goes without saying that the study of universalities is, in and of itself, an extremely interesting and fertile pursuit. Add to this our current proximity to the release of LHC results and the requisite grasp of strong coupling gauge dynamics and it becomes evident that a more complete glossary of the quantities that are in the same universality class in large $N$, $\mathcal{N}=4$ SYM and QCD is crucial. We hope that the arguments presented here will, if nothing else, stimulate further research in this direction.
\section{Acknowledgments}
It is with much gratitude that we acknowledge Robert de Mello Koch, Brian Greene, Antal Jevicki, Robert McNees and Horatiu Nastase for stimulating discussion. J.M. and A.H. are supported by NRF Thuthuka grant GUN61699 and A.P. by an NRF Scarce Skills PhD fellowship.  



\begin{thebibliography}{99}
\bibitem{Maldacena97}
J. M. Maldacena, 
``The Large N limit of superconformal field theories and supergravity"
Adv. Theor. Math. Phys. {\bf 2} 231 (1998)
\texttt{hep-th/9711200};
 
\bibitem{Alday-Maldacena07}
L. Alday and J. Maldacena,
``Gluon scattering amplitudes at strong coupling,"
\texttt{hep-th/0705.0303}.
 
\bibitem{PSS01}
G. Policastro, D.T. Son and A.O. Starinets
``The Shear viscosity of strongly coupled N=4 supersymmetric Yang-Mills plasma,"
  Phys.\ Rev.\ D {\bf 87}, 081601 (2001)
\texttt{hep-th/0104066}.

\bibitem{Polchinski95}
J. Polchinski
``Dirichlet Branes and Ramond-Ramond charges,"
Phys.\ Rev.\ Lett. {\bf 75}, 4724-4727 (1995)
\texttt{hep-th/9510017}

\bibitem{MST00}
J. McGreevy, L. Susskind and N. Toumbas 
``Invasion of the giant gravitons from Anti-de Sitter space,"  
 JHEP {\bf 0006}, 008 (2000)
\texttt{hep-th/0003075}

\bibitem{Hagedorn-Refs}
E.Alvarez,
``Strings At Finite Temperature",  
Nucl. Phys. B {\bf 269}, 596 (1986)

E.Alvarez and M.A. Osorio,
``Superstrings at Finite Temperature,"
Phys.\ Rev.\ Lett. {\bf 36}, 1175 (1987)

J.J. Atcik and E. Witten,
``The Hagedorn Transition and the Number of Degrees of Freedom of String Theory",
Nucl. Phys. B {\bf 310}, 291 (1988)

\bibitem{SYM Hagedorn refs}
 E. Witten,
 ``Anti-de Sitter Space, Thermal Phase Transition, and Confinement in Gauge Theory"
 Adv. Theor. Math. Phys. {\bf 2}, 505 (1998)

  B.~Sundborg,
  ``The Hagedorn transition, deconfinement and N = 4 SYM theory,''
  Nucl. Phys. B {\bf 573}, 349 (2000)
  \texttt{hep-th/9908001}.
  
\bibitem{BMN}
  D. Berenstein, J.M. Maldacena and H. Nastase,
  ``Strings in flat space and pp waves from $\mathcal{N} = 4$ Super Yang Mills,''
  JHEP {\bf 0204}, 013 (2002)
  \texttt{hep-th/0202021}.
 
\bibitem{Metsaev:2001bj}
  R.R. Metsaev,
  ``Type IIB Green-Schwarz superstring in plane wave Ramond-Ramond
  background,''
  Nucl. Phys. B {\bf 625}, 70 (2002)
  \texttt{hep-th/0112044}.

  R.R. Metsaev and A.A. Tseytlin,
  ``Exactly solvable model of superstring in plane wave Ramond-Ramond
  background,''
  Phys.\ Rev.\ D {\bf 65}, 126004 (2002)
  \texttt{hep-th/0202109}.
   
\bibitem{pp-wave Hagedorn} 
  L.A. Pando Zayas and D. Vaman,
  ``Strings in RR plane wave background at finite temperature,"
  Phys. Rev. {\bf D67}, 106006 (2003)
  \texttt{hep-th/0208066}
  
  B.R. Greene, K. Schalm and G. Shiu,
   ``On the Hagedorn behaviour of pp-wave strings and N = 4 SYM theory at finite R-charge density,''
  Nucl. Phys. B {\bf 652}, 105 (2003)
  \texttt{hep-th/0208163}.

  R.C. Brower, D.A. Lowe and C.I. Tan,
   ``Hagedorn transition for strings on pp-waves and tori with chemical potentials,''
  Nucl. Phys. B {\bf 652}, 127 (2003)
  \texttt{hep-th/0211201}.
  
  G. Grignani, M. Orselli, G.W. Semenoff and D. Trancanelli,
  ``The Superstring Hagedorn temperature in a pp wave background,"
  JHEP {\bf 06}, 006 (2003)
  \texttt{hep-th/0301186}
  
  F. Bigazzi and A.L. Cotrone,
  ``On zero point energy, stability and Hagedorn behavior of type IIB strings on pp waves,"
  JHEP {\bf 08}, 052 (2003)

Y.~Sugawara,
``Thermal amplitudes in DLCQ superstrings on
pp-waves,''
Nucl.\ Phys.\ B {\bf 650}, 75 (2003)
[arXiv:hep-th/0209145].

Y.~Sugawara,
``Thermal partition function of superstring on
compactified pp-wave,''
Nucl.\ Phys.\ B {\bf 661}, 191 (2003)
[arXiv:hep-th/0301035].

 M.~Blau, M.~Borunda and M.~O'Loughlin,
  ``On the Hagedorn behaviour of singular scale-invariant plane waves,''
  JHEP {\bf 0510}, 047 (2005)
  [arXiv:hep-th/0412228].
  
  \bibitem{Matching Hagedorns}
  T. Harmark and M. Oreselli, 
  ``Quantum Mechanical Sectors in Thermal $\mathcal{N}=4$ Super Yang-Mills on 
  $\mathbb{R}\times S^{3}$,'' 
  Nucl. Phys. B \textbf{757}, 117-145 (2006) 
  \texttt{hep-th/0605234}.
  
  T. Harmark and M. Orselli 
  ``Matching the Hagedorn temperature in AdS/CFT"
  Phys.\ Rev.\ D {\bf 74}, 126009 (2006)
  \texttt{hep-th/0608115}.
  
  T. Harmark, K. Kristj\'{a}nsson and M. Oreselli, 
  ``Decoupling Limits of $\mathcal{N}=4$ super Yang-Mills on $\mathbb{R}\times S^{3}$" 
  \texttt{hep-th/0707.1621}.

\bibitem{LM}
O. Lunin and J. Maldacena, 
``Deformed field theories with $U(1)\times U(1)$ global symmetry and their gravity duals,'' 
JHEP \textbf{0505}, 033 (2005) 
\texttt{hep-th/0502086}.  

\bibitem{Leigh-Strassler}
R. G. Leigh and M. J. Strassler
``Exactly marginal operators and duality in four-dimensional N=1 supersymmetric gauge theory,"
Nucl.Phys. {\bf B447}, 95-136 (1995)
\texttt{hep-th/9503121}

\bibitem{Niarchos-Prezas}
V. Niarchos and N. Prezas, 
``BMN operators for $\mathcal{N}=1$ Superconformal Yang-Mills Theories and Associated String Backgrounds,'' JHEP \textbf{0306}, 015 (2005) 
\texttt{hep-th/0212111}

\bibitem{Hpp}
M. Blau, J. Figueroa-O'Farrill, C. Hull and G. Papadopoulos,
``A New maximally supersymmetric background of IIB superstring theory,"
JHEP \textbf{0201}, 047 (2002)
 \texttt{hep-th/0110242}.

Matthias Blau and M. O'Loughlin, 
``Homogeneous plane waves,"
 Nucl.Phys. {\bf B654}, 135-176 (2003)
\texttt{hep-th/0212135}

M. Blau, M. O'Loughlin, G. Papadopoulos and A. A. Tseytlin,
``Solvable models of strings in homogeneous plane wave backgrounds,"
 Nucl.Phys. {\bf B673}, 57-97 (2003)
\texttt{hep-th/0304198}

\bibitem{DMSS}
R. de Mello Koch, J. Murugan, J. Smolic and M. Smolic, 
``Deformed PP-waves from the Lunin-Maldacena Background,''
 JHEP \textbf{0508}, 072 (2005) 
 \texttt{hep-th/0505227}.
 
\bibitem{Mateos}
Toni Mateos,
``Marginal deformation of ${\cal N}=4$ SYM and Penrose limits with continuum spectrum",
J. High Energy Phys {\bf 0508}, 026, (2005), 
\texttt{hep-th/0505243} 

\bibitem{Deformed giants}
A. Hamilton and J. Murugan
``Giant Gravitons on Deformed pp-waves,"
J. High Energy Phys {\bf 0706}, 036, (2007), 
\texttt{hep-th/0609135}

M. Pirrone
``Giants on Deformed Backgrounds,"
J. High Energy Phys {\bf 0612}, 064, (2006), 
\texttt{hep-th/0609173} 

\bibitem{Bigazzi}
F. Bigazzi and A.L. Cotrone,
``On zero point energy, stability and Hagedorn behavior of type IIB strings on pp waves"
J. High Energy Phys {\bf 0308}, 052, (2003), 
\texttt{hep-th/0306102} 
  
\bibitem{Takayanagi:2002pi}
  T. Takayanagi,
  ``Modular invariance of strings on pp-waves with RR-flux,''
  JHEP {\bf 0212}, 022 (2002)
  \texttt{hep-th/0206010}.
 
\bibitem{gradrhyz}
I.\ Gradshteyn and I.\ Ryzhik,
``Table of integrals series and products'', sixth edition,
Academic Press, (2000).

\bibitem{plefka}
J. Plefka, 
``Lectures on the Plane-Wave String/Gauge Theory Duality,'' 
Fortsch. Phys. \textbf{52}, 264-301 (2004)
\texttt{hep-th/0307101}.

\bibitem{deformed-spin-chain}
D. Berenstein and S.A. Cherkis,
``Deformations of $\mathcal{N}=4$ SYM and integrable spin-chain models",
Nucl. Phys. {\bf B 702}, 49 (2004)
\texttt{hep-th/0405215}

N. Beisert and R. Roiban,
``Beauty and the twist",
JHEP {\bf 0508}, 039 (2005)
  \texttt{hep-th/0505187}.
  
S.A. Frolov, R. Roiban and A.A. Tseytlin,
``Gauge-string duality for superconformal deformations of $\mathcal{N}=4$ super Yang-Mills theory,"
JHEP {\bf 0507}, 045 (2005)
\texttt{hep-th/0503192}  

\bibitem{Gursoy}
U Gursoy,
"Probing universality in the gravity duals of $\mathcal{N}=1$ SYM by gamma-deformations",
JHEP {\bf 0605}, 014 (2006)
\texttt{hep-th/0602215} 

\bibitem{MN}
J.M. Maldacena and C. Nu$\tilde{\rm n}$ez,
"Towards the large $N$ limit of pure $\mathcal{N}=1$ superYang-Mills",
Phys.Rev.Lett. {\bf 86}, 588-591 (2001)
\texttt{hep-th/0008001}

\bibitem{FRT}
S.A. Frolov, R. Roiban and A.A. Tseytlin,
``Gauge-string duality for (non)supersymmetric deformations of 
$\mathcal{N}=4$ super Yang-Mills theory,"
Nucl. Phys. {\bf B 731}, 1-44 (2005)
\texttt{hep-th/0507021}

\bibitem{Frolov}
S.A. Frolov,
``Lax pair for strings in Lunin-Maldacena background,"
JHEP {\bf 0505}, 069 (2005)
\texttt{hep-th/0503201} 

\bibitem{Swanson}
T. McLoughlin and I. Swanson,
``Integrable twists in AdS/CFT,"
JHEP {\bf 0608}, 084 (2006)
\texttt{hep-th/0605018}

\bibitem{KS}
I.R. Klebanov and M.J. Strassler,
`` Supergravity and a confining gauge theory: Duality cascades and $\chi$ SB resolution of naked singularities,"
JHEP {\bf 0008}, 052 (2000)
\texttt{hep-th/0007191}

\end{thebibliography}
 \end{document}